\documentclass{PoS}

\newcommand{\be}{\begin{equation}} \newcommand{\ee}{\end{equation}}
\newcommand{\beq}{\begin{equation}} \newcommand{\eeq}{\end{equation}}
\newcommand{\beqa}{\begin{eqnarray}}
\newcommand{\eeqa}{\end{eqnarray}} \newcommand{\eq}[1]{(\ref{#1})}
\def\nn{\nonumber} \def\bea{\begin{eqnarray}} \def\eea{\end{eqnarray}}

\newcommand{\barr}{\begin{array}}
\newcommand{\earr}{\end{array}}

\def\a{\alpha}  
 \def\g{\gamma} \def\G{\Gamma}
 \def\d{\delta} 
 \def\e{\epsilon} 
\def\f{\phi}    
\def\l{\lambda}   \def\m{\mu}
\def\n{\nu} \def\o{\omega}   \def\r{\rho}
\def\s{\sigma}   \def\th{\theta}
   \def\z{\zeta}

     \def\cM{{\cal M}} \def\cN{{\cal N}}

\def\R{{\mathbb R}} \def\C{{\mathbb C}} 
\def\Z{{\mathbb Z}} \def\one{\mbox{1 \kern-.59em {\rm l}}}

\def\bit{\begin{itemize}} \def\eit{\end{itemize}} 

\def\({\left(} \def\){\right)}

\title{Orbifold matrix models and fuzzy extra dimensions}

\ShortTitle{Orbifold matrix models}

\author{\speaker{Athanasios Chatzistavrakidis}
\\
        Bethe Center for Theoretical Physics and Physikalisches Institut, University of Bonn\\
Nussallee 12, D-53115 Bonn, Germany\\
        E-mail: \email{than@th.physik.uni-bonn.de}}

\author{Harold Steinacker\\
      Faculty of Physics, University of Vienna\\
Boltzmanngasse 5, A-1090 Vienna, Austria\\ and \\
{\it  Physics Department\\
City College of the City University of New York\\
160 Convent Avenue, New York, NY 10031 }\\
       E-mail: \email{harold.steinacker@univie.ac.at}}

\author{George Zoupanos 
\thanks{On leave from Physics Department, National Technical University, Zografou Campus:
Heroon Polytechniou 9, 15780 Zografou, Athens, Greece} \\
      Theory Group, Physics Department \\
CERN, Geneva, Switzerland\\
       E-mail: \email{George.Zoupanos@cern.ch}}

\abstract{We revisit an orbifold matrix model obtained as a restriction of the type IIB matrix model on a $\Z_3$-invariant sector. 
An investigation of its moduli space of vacua is performed 
and issues related to chiral gauge theory and gravity are discussed. Modifications of the orbifolded model triggered 
by Chern-Simons or mass deformations are also analyzed. Certain vacua of the modified models 
exhibit higher-dimensional behaviour with internal geometries related 
to fuzzy spheres.}

\FullConference{Proceedings of the Corfu Summer Institute 2011 School and Workshops on Elementary Particle Physics and Gravity\\
		September 4-18, 2011\\
		Corfu, Greece
\\ \vspace{10pt}
Preprint number: CERN-PH-TH/2012-057
}

\begin{document}

\section{Introduction}

The theoretical efforts to establish a deeper understanding of Nature has
led to a huge and sustained research activity in two very interesting
frameworks, namely Superstring Theories \cite{Green:1987sp} and Non-Commutative Geometry
\cite{ncg}. Superstring Theory is often regarded as the best candidate for a
quantum theory of gravitation, or more generally as a unified theory of
all fundamental interactions. Similarly Non-Commutative space-time and in
general Non-Commutative Geometry is expected, on quite general grounds, to
be the natural arena to study a quantum theory of gravity. In addition the
two frameworks came closer by realizing that in M-Theory and in open
String Theory, in the presence of a non-vanishing background antisymmetric
field, the effective physics on D-branes can be described by a
Non-Commutative gauge theory \cite{Connes:1997cr,Seiberg:1999vs}. Thus Non-Commutative field theories
emerge as effective description of string dynamics. Moreover the type IIB
superstring theory, related to the other superstring theories by certain
dualities, in its conjectured non-perturbative formulation as a matrix
model \cite{Ishibashi:1996xs}, is a non-commutative theory. Finally the String Theory and
Non-Commutative Geometry approaches share similar hopes for exhibiting
improved renormalization properties in the ultraviolet regime as compared
to ordinary field theories.

Non-commutative geometry has been regarded as a promising framework for
constructing finite quantum field theories or at least as a natural scheme
for regularizing quantum field theories. However the quantization of field
theories on non-commutative spaces has turned out to be much more
difficult than expected and with problematic ultraviolet features \cite{Filk:1996dm}, see
however \cite{Grosse:2004ik}, and \cite{Steinacker:2003sd}. Although models of the Standard Model (SM) type have been constructed
using the Seiberg-Witten map, they can only be considered as effective
theories and they are not renormalizable. A drastic change in the
perspective of non-commutative geometry was given with the suggestion that
indeed it might be relevant for particle physics models but in the
description of extra dimensions \cite{Aschieri:2003vy}. The higher-dimensional theories that
can be constructed based on this proposal, appeared to have many
interesting unexpected features ranging from their renormalizability to
their potential predictivity.

In the context of higher-dimensional gauge theories, we have explored the
possibility that the extra dimensions are described by matrix
approximations to smooth manifolds known as fuzzy spaces. Initially,
higher-dimensional gauge theories defined on the product of Minkowski
space and a fuzzy coset space $(S/R)_F$ were considered and their
dimensional reduction to four dimensions was performed, using the Coset
Space Dimensional Reduction scheme (CSDR) \cite{Forgacs:1979zs,Kapetanakis:1992hf,Kubyshin:1989vd}. 
Although for the
technicalities one has to consult the original or some review papers \cite{Aschieri:2003vy},
a major difference between fuzzy and ordinary CSDR is that the
four-dimensional gauge group appearing in the ordinary CSDR after the
geometrical breaking and before the spontaneous symmetry breaking due to
the four-dimensional Higgs fields does not appear in the fuzzy CSDR. In
fuzzy CSDR the spontaneous symmetry breaking takes already place by
solving the fuzzy CSDR constraints and the four-dimensional potential
appears already shifted to a minimum. Therefore in four dimensions appears
only the physical Higgs field that survives after a spontaneous symmetry
breaking. Correspondingly in the Yukawa sector of the theory we have the
results of the spontaneous symmetry breaking, i.e. massive fermions and
Yukawa interactions among fermions and the physical Higgs field. The
conclusion is that if one would like to describe the spontaneous symmetry
breaking of the SM in this framework, then one would be naturally led to
consider large extra dimensions. A fundamental difference between the
ordinary CSDR and its fuzzy version is the fact that a non-Abelian gauge
group G is not really required in many dimensions. Indeed it turns out
that the presence of a $U(1)$ in the higher-dimensional theory is enough to
obtain non-Abelian gauge theories in four dimensions.

Another fundamental difference as compared to all known theories defined
in more than four dimensions is that the present ones are renormalizable.
Arguments were given supporting this observation \cite{Aschieri:2003vy} but the strongest
support came by studying the whole problem from another point of view.
Specifically, in a further important development we presented a
renormalizable four-dimensional $SU(N)$ gauge theory with a suitable
multiplet of scalars, which dynamically develops fuzzy extra dimensions
that form a fuzzy sphere \cite{Aschieri:2006uw}. 
This model has non-trivial vacua which admit an interpretation 
 as 6-dimensional gauge theory, 
with gauge group and geometry depending on the
parameters appearing in the original Lagrangian. 
A finite  tower of massive Kaluza-Klein modes arises, consistent with an
interpretation as compactified higher-dimensional gauge theory. 
There are many
remarkable aspects of this model. First, it provides an extremely simple
and geometrical mechanism of dynamically generating extra dimensions. This is
based on an important lesson from non-commutative gauge theory,
namely that non-commutative or fuzzy spaces can be obtained as solutions
of matrix models. The mechanism is quite generic, and does not require
fine-tuning or supersymmetry. This provides in particular a realization of
the basic ideas of compactification and dimensional reduction within the
framework of renormalizable quantum field theory. Moreover, being essentially
a large $N$ gauge theory,
the analytical techniques developed in this context 
should be applicable. In particular,
it turns out that the generic low-energy gauge group is given by $SU(n_1) \times
SU(n_2) \times U(1)$ or $SU(n)$, while gauge groups which are products of more
than two simple components (apart from $U(1)$) do not seem to occur in this
model. Moreover, a magnetic flux is naturally induced in the vacua with
non-simple gauge group.

The appealing features of the above mechanism naturally suggest to find models which are 
phenomenologically relevant to particle
physics. One of the most challenging problems in this direction is
the accommodation of chiral fermions in four dimensions. The inclusion of
fermions in the above class of models showed that the best one could
achieve without further requirements is to obtain theories with mirror
fermions in bi-fundamental representations of the low energy gauge group
\cite{Steinacker:2007ay}. Indeed, studying in detail the fermionic sector of models which
dynamically develop extra dimensions with the geometry of a fuzzy sphere
or a product of two fuzzy spheres we found  \cite{Chatzistavrakidis:2009ix} that the low-energy theory
contains two mirror sectors, even when magnetic fluxes are included on the
two fuzzy spheres. Although mirror fermions do not exclude the possibility
to make contact with phenomenology \cite{Maalampi:1988va}, it is clearly desirable to obtain
exactly chiral fermions. This was achieved by introducing an additional
structure in the above context, based on orbifolds. In particular,
a $\Z_3$ orbifold projection of a $N=4\ \  SU(3N)$ SYM theory leads to a
$N=1$ supersymmetric theory with gauge group $SU(N)^3$ \cite{Chatzistavrakidis:2010xi}. Adding a suitable set of
soft supersymmetry breaking terms in the $N=1$ theory, certain vacua of the
theory were revealed, where twisted fuzzy spheres are dynamically
generated. It is well known that the introduction of a soft supersymmetry
breaking sector is not only natural but also necessary in the
constructions of phenomenologically viable supersymmetric theories, with
prime example the case of the MSSM. Such vacua correspond to models which
retain the welcome features described above and in addition they exhibit a
chiral low-energy spectrum. The potentially most interesting chiral models
for low-energy phenomenology which can be constructed in this context turn
out to be $SU(4) \times SU(2) \times SU(2)$, $SU(4)^3$ and $SU(3)^3$. Eventually the most
interesting case turned out to be the a unified theory based on the
trinification group, which can be promoted even to an all-loop finite
theory (for a review see \cite{Heinemeyer:2010xt}) and moreover it is suitable to make
testable predictions \cite{Ma:2004mi}. Thus it was shown that fuzzy extra dimensions
can arise in simple field-theoretical models which are chiral,
renormalizable, and maybe phenomenologically viable.

In the present contribution, we would like to report on a natural extension of the above
ideas and methods, realized in the context of  Matrix Models (MM). MMs offer a framework where both
profound conceptual problems as well as questions about low-energy physics
may be addressed. Indeed, at a fundamental level, the MMs introduced by
Banks-Fischler-Shenker-Susskind (BFSS) and
Ishibashi-Kawai-Kitazawa-Tsuchiya (IKKT) are supposed to provide a
non-perturbative definition of M theory and type IIB string theory
respectively \cite{Ishibashi:1996xs,Banks:1996vh}. On the other hand, MMs are also useful laboratories
for the study of structures which could be relevant from a low-energy
point of view. Indeed, they generate a plethora of interesting solutions,
corresponding to strings, D-branes and their interactions \cite{Ishibashi:1996xs,Chepelev:1997ug}, as well
as to non-commutative/fuzzy spaces, such as fuzzy tori and spheres
\cite{Iso:2001mg}. Such backgrounds naturally give rise to non-abelian gauge
theories. Therefore it appears natural to pose the question whether it is
possible to construct phenomenologically interesting particle physics
models in this framework as well. 
%This was the main objective of our
%recent work initiated in [23].

In particular, MMs also allow to explore
ideas and lessons which initially arose in the context of orbifolds. An
orbifold MM was proposed by Aoki-Iso-Suyama (AIS)  in \cite{Aoki:2002jt} as a particular projection of the IKKT
model, and it is directly related to our construction in which fuzzy extra
dimensions arise with trinification gauge theory \cite{Chatzistavrakidis:2010xi}. By $\Z_3$-orbifolding,
the original symmetry of the IKKT matrix model with matrix size $3N \times 3N$ is
generally reduced from $SO(9,1) \times U(3N)$ to $SO(3,1) \times U(N)^3$. This model is
chiral and has $D=4, \ \  N=1$ supersymmetry of Yang-Mills type as well as an
inhomogeneous supersymmetry specific to matrix models. The $\Z_3$-invariant
fermion fields transform as bi-fundamental representations under the
unbroken gauge symmetry exactly as in our constructions.

In the following we present first the field theoretic construction
described above, which we then attempt to embed in the existing
Orbifold MM \cite{Aoki:2002jt} performing the appropriate modifications. We briefly
speculate on the gravity that emerges out of this picture, 
and we leave for future work the phenomenological consequences resulting from such an embedding, 
noting that the four-dimensional non-commutativity appears as a new feature.

\section{Field theory orbifolds and fuzzy spheres}
\label{sec:chiral-gauge}

Let us begin with a brief reminder of the application of orbifold techniques in field theory and their 
relation to the dynamical generation of fuzzy extra dimensions. The main motivation behind this concerns 
the possibility to obtain chiral matter when models for particle physics are engineered.

The starting point is a $SU(3N)$ ${\cal N}=4$ supersymmetric Yang-Mills (SYM) theory.
This theory may be projected under the action of the discrete group $\Z_3$; this is essentially 
an orbifold projection. This procedure amounts to an embedding of the discrete symmetry into the $R$-symmetry 
of the original theory, which is $SU(4)_R$. It is well-known that depending on this embedding, projected 
theories with different amount of remnant supersymmetry may be obtained \cite{Kachru:1998ys}. Indeed, 
a maximal embedding $\G\in SU(4)_R$ breaks supersymmetry completely, while embeddings within an $SU(3)$ or 
$SU(2)$ part of the $R$-symmetry lead to ${\cal N}=1$ or ${\cal N}=2$ supersymmetric theories respectively.
In the present work our attention is focused on the first of the latter two possibilities, which is in accord 
with our motivation of obtaining chiral theories.

A projection of the initial theory under the discrete symmetry $\Z_3$ corresponds to an ${\cal N}=1$ SYM 
theory where only the $\Z_3$-invariant fields survive. The details of this procedure in the present context 
may be found in \cite{Chatzistavrakidis:2010xi}; here we just list the most important results. The original ${\cal N}=4$ SYM theory 
contains a vector superfield and three chiral superfields in ${\cal N}=1$ language. The component fields of the theory are the 
gauge fields $A_{\m}, \m=0,\dots 3$ of the $SU(3N)$ gauge group, three complex scalars $\phi^i, i=1,2,3$, 
trnasforming in the adjoint of the gauge group and in the vector of the global symmetry, and four 
Majorana fermions $\psi^p$, in the adjoint of the gauge group and the spinor of the global symmetry. 
After the projection, the resulting theory has a different gauge group and spectrum. Indeed, the gauge group 
is broken down to $H=S(U(N)\times U(N)\times U(N))$, while the surving scalar fields and fermions transform 
according to the representations
\be
\label{repsH} 3\cdot \biggl((N,\overline{N},1)+(\overline{N},1,N)+(1,N,\overline{N})\biggl)
\ee 
under the non-abelian gauge group factors. This spectrum is evidently free of non-abelian gauge anomalies.
Moreover, it is directly observed that the matter fields transform in chiral representations of the gauge group and they are replicated three times, 
thus leading to three chiral families. 

Furthermore, the F-term scalar potential of the $\cN=4$ SYM theory is
\be 
V_F(\f)= \frac 14 \mbox{Tr}([\f^i,\f^j]^\dagger [\f^i,\f^j]), 
\ee 
and it formally remains the same after the projection with the obvious difference that it describes 
only the interactions among the surviving fields.
The D-term contribution to the scalar potential is given by $V_{D}=\frac 12 D^2=\frac 12 D^{I}D_{I}$, 
where the D-terms have the form $D^{I}=\f_i^{\dagger}T^{I}\f^i$, with $T^I$ the generators 
of the representation of the corresponding chiral multiplets.  The minimum of the full potential is obtained 
for vanishing F- and D-terms, which means that $[\f^i,\f^j]=0$. However, a very different class of vacua 
may be revealed upon adding a soft supersymmetry breaking sector in the theory. The scalar part of this 
sector is
\be\label{soft} 
V_{SSB}=\frac 12 \sum_i m^2_i\, {\f^i}^{\dagger}\f^i+\frac 12 \sum_{i,j,k}\, h_{ijk}\f^i\f^j\f^k+h.c., 
\ee
and it respects the orbifold symmetry.
 After the addition of these soft terms the 
scalar potential of the theory becomes 
\be\label{potential1} 
V=V_F+V_{SSB}+V_{D},
\ee
which has the equivalent form
\be\label{potnewform} 
V=\frac 14 (F^{ij})^{\dag}F^{ij} \,\, + V_D 
\ee
for suitable parameters, where we have defined
\be
\label{fieldstrength} 
F^{ij}=[\f^i,\f^j]-i\e^{ijk}(\f^k)^{\dagger}.
\ee
Since the first term is positive definite, the global minimum of the potential is obtained when the following relations hold,
\bea
\label{twisted-vacuum} 
[\f^i,\f^j]&=&i\e_{ijk}(\f^k)^{\dagger}, \\ 
{[}(\f^i)^{\dagger},(\f^j)^{\dagger}]&=&i\e_{ijk}\f^k, \\
\f^i (\f^i)^{\dagger} &=&  R^2,
\eea
where $(\f^i)^{\dagger}$ denotes hermitean conjugation of the complex scalar field $\f^i$
and $[ R^2,\f^i] = 0$.
The above relations are closely related to the usual fuzzy sphere.
This can be seen by considering the untwisted fields $\tilde \f_i$, defined by
\be
\f^i = \Omega\,\tilde\f^i , 
\label{twisted-fields}
\ee
for some $\Omega\ne 1$ which satisfies
$
\Omega^3 = 1,  [\Omega,\f^i] = 0, \Omega^\dagger = \Omega^{-1}
\label{cond-1}
$
and\footnote{This is understood before the orbifolding.}
$
(\tilde\f^i)^\dagger = \tilde\f^i, \mbox{i.e.}\,\,(\f^i)^\dagger = \Omega \f^i.
\label{cond-2}
$
Then \eq{twisted-vacuum} reduces to the ordinary fuzzy sphere relation
\be
\,[\tilde\f^i,\tilde\f^j] = i\e_{ijk}\tilde\f^k,
\label{fuzzy-transf} 
\ee
generated by $\tilde\f^i$, as well as to the relation
$
\tilde \f^i\tilde \f^i = R^2.
$

Configurations of $\f^i$ satisfying the relations (\ref{twisted-vacuum}) have the form
\be
\label{solution1} 
\phi^i = \Omega\, (\one_3\otimes\lambda^i_{(N)}),
\ee 
where $\l^i_{(N)}$ denote the generators of $SU(2)$ in the $N$-dimensional irreducible representation and the matrix $\Omega$ is given by
\be
\Omega = U \otimes \one_N, \quad
U = \left(\begin{array}{ccc}
		0 & 1 & 0 \\
		0 & 0 & 1 \\
		1 & 0 & 0 \\
\end{array}\right), \quad U^3 = \one. 
\label{omega-large}
\ee 
This solution completely breaks the gauge symmetry $SU(N)^3$ (one could say that it corresponds to the Higgs 
branch of the SYM theory). 
However, there exist solutions which do not break the gauge symmetry completely, such as
\be
\phi^i = \Omega\, \biggl(\one_3\otimes(\lambda^i_{(N-n)} \oplus 0_{n})\biggl),
\label{twistedfuzzys2-2}
\ee
where $0_{n}$ denotes the $n\times n$ matrix with vanishing entries. 
The gauge symmetry in this case is broken from $SU(N)^3$ down to $SU(n)^3$ and the vacuum is interpreted as $\R^4 \times K_F$
with an internal fuzzy geometry $K_F$ of a twisted fuzzy sphere.
Such vacua may lead to a phenomenologically interesting low-energy theory, as discussed in \cite{Chatzistavrakidis:2010xi}.

\section{Orbifold matrix models}

The IKKT matrix model was originally proposed in \cite{Ishibashi:1996xs} as a non-perturbative definition of the type IIB superstring theory. It is a reduced matrix model defined by the following action functional,
\be
{\mathcal S}_{IKKT}  =  -{1\over 2g^2}\mbox{Tr}\biggl({1\over 2}[X_a,X_b][X^a,X^b]
+\bar{\Psi}\Gamma ^a[X_a,\Psi ]\biggl),
\label{IKKT-action}
\ee
where $X^a, ~a=0,\dots,9$ are ten hermitian matrices, $\Psi$ are 16-component Majorana-Weyl spinors and $g$ is a coupling parameter. Moreover, $\Gamma^a$ are 10-dimensional matrices furnishing the spinor representation of $SO(10)$, which is a symmetry of the Euclidean model inherited by its 10-dimensional origin. 
The full symmetry group of the above model contains the $U(N)$ gauge group
(where actually $N \to \infty$ is understood) as well as the aforementioned $SO(10)$ global symmetry
and a ${\cal N}=2$ supersymmetry.
The action (\ref{IKKT-action}) may be varied with respect to the matrices $X_a$ and, setting $\Psi=0$, the resulting equations of motion are,
\be\label{ikkteoms} [X_b,[X^a,X^b]]=0. \ee 
Classical solutions of the model can be obtained in the case when the matrices are commuting, $[X^a,X^b]=0$,
but also when they do not commute, $[X^a,X^b]\ne0$; in the rest of this work we focus on the second possibility since the former one does not generate interesting dynamics. Furthermore, since we are going to describe non-commutative (NC) branes embedded in $\R^{10}$, 
let us consider a splitting of the ten matrices $X^a$ as  
\be\label{split} X^a=  \left(\begin{array}{c}X^\mu \\
           X^{i}
             \end{array}\right),\quad \m=0,\dots,2d, \quad i=2d+1,\dots,9. \ee 
Then the $X^{\m}$ correspond to the world-volume directions on the branes and the $X^i$ correspond to the 
transverse displacements.

 In the above notation, a solution of the equations (\ref{ikkteoms}) is given by
\be
 X^a = \left(\begin{array}{c}\bar X^\mu \\
          0
             \end{array}\right) \ee 
where $\bar X^{\m}$ are the generators of the Moyal-Weyl quantum plane $\R^{2d}_{\theta}$, which satisfy the commutation relation
\be [\bar X^\mu,\bar X^\nu] = i \theta^{\mu\nu},
\ee
where $\theta^{\m\n}$ is a constant antisymmetric tensor. 
This solution preserves half of the supersymmetries of the IKKT model, and corresponds to a single non-commutative flat $(2d-1)$-brane. 
Being a single brane, this solution is associated to an abelian gauge theory. 
Non-abelian gauge theories are implemented via direct generalization of the above solution, 
\be
 X^a = \left(\begin{array}{c}\bar X^\mu \\
          0
             \end{array}\right)\otimes \one_{n},
\ee
where $\one_n$ is the $n$-dimensional unit matrix. 
This configuration is interpreted as $n$ coincident branes carrying a non-abelian $U(n)$ gauge theory, 
similarly to ordinary D-branes \cite{Witten:1995im}. 

The flat solutions $\R^{2d}_\theta$ are special cases of NC branes with
generic embedding in $\R^{10}$. Such generic branes are desribed by 
 quantized embedding functions
$$
X^a \sim x^a: \quad \mathcal{M}^{2n}\hookrightarrow \R^{10} 
$$ 
of a $2n$ dimensional submanifold. Furthermore,
$$
[X^\mu,X^\nu] \sim  i \{x^\mu,x^\nu\} = i \theta^{\mu\nu}(x)\,
$$
is interpreted as a quantized Poisson structure on $\mathcal{M}^{2n}$.
Here $\sim$ denotes the
semi-classical limit where commutators are replaced by Poisson brackets, and
$x^\mu$ are locally independent coordinate functions chosen among the $x^a$.
Thus we are considering quantized embedded
Poisson manifolds $(\cM^{2n},\theta^{\mu\nu})$.
The sub-manifold $\cM^{2n}\subset\R^{10}$
is equipped with a non-trivial induced metric
$$
g_{\mu\nu}(x)=\partial_\mu x^a \partial_\nu x^b\eta_{ab}\,,
\label{eq:def-induced-metric}
$$
via the pull-back of $\eta_{ab}$.
 The kinetic term for
all (scalar, gauge and fermionic) fields in the MM on such a background $\cM^{2n}$
is governed (up to possible conformal factors) by the effective metric
\be
G^{\m\n} \sim \th^{\m\r}\th^{\n\s}g_{\r\s}\,,
\label{eff-metric}
\ee
so that $G_{\mu\nu}$ must be interpreted as gravitational metric.
Since the embedding is dynamical, the model describes a dynamical
theory of gravity, realized on dynamically
determined submanifolds of $\R^{10}$.

The IKKT matrix model is very interesting in view of its conjectured merit to capture non-perturbative type IIB superstring theory. 
However, it is not obvious how one could relate it directly to the observed four-dimensional world and make contact with low-energy phenomenology. In particular, its main limitation is the difficulty in the description of chiral fermions, since it accommodates matter in real representations of the gauge group.
The above problem may be addressed in a variety of ways. Here, motivated by the 
discussion of the previous section,
 we discuss the possibility 
to construct orbifold equivalents within the framework of matrix models. Such an approach was initially 
employed by Aoki, Iso and Suyama in \cite{Aoki:2002jt}.
Orbifold techniques in bottom-up stringy constructions, i.e. from a 4-dimensional perspective, were analyzed in \cite{Aldazabal:2000sa} in the context of phenomenologically attractive model building with D-branes. 
Here we find it important to retain the 0-dimensional perspective of reduced matrix models and study orbifold techniques in this context.  

In order to apply the aforementioned orbifold techniques in our case, we have to specify the embedding of some discrete symmetry in the full symmetry group $U(N)\times SO(10)$ of the IKKT model. The basic steps 
closely follow the corresponding discussion for field theories which was sketched in section 2. In the present context 
 we discuss the discrete 
group $\Z_3$ too. Clearly there is a huge number of different embeddings which can be considered. However, in accordance with the lessons from string theory compactifications and field-theoretic orbifolds of ${\cal N}=4$ SYM theory \cite{Douglas:1997de,Kachru:1998ys}, 
it is reasonable to embed $\Z_3$ in the $SO(6)$ part of the global $SO(10)$ symmetry. 

According to the above, let us choose six out of the nine space-like matrices of the IKKT model
and write them in complex form
$Z_i,~i=1,2,3$,
\be
Z_1=X_4 + iX_5, \ \ \ Z_2=X_6 + iX_7, \ \ \ Z_3=X_8 + iX_9.
\ee
Then the action of ${\Z}_3$ on the matrices is characterized by three
integers $(a_1,a_2,a_3)$, one for each complex matrix \cite{Douglas:1997de}. 
Let us denote with
\be
\zeta \in \Z_3
\ee
the abstract generator of $\Z_3$.
The ${\Z}_3$ transformation is
defined to act on the above complex matrices as
\be
\zeta ( Z_i) = \omega^{a_i} Z_i.
\ee
On the other hand, its action on the rest four of the ten matrices is trivial, i.e.
\be \z( X^{\m})=X^{\m}. \ee
The important point here is that the realization of the $\Z_3$ action on the indices, i.e. on $\R^{10}$, is chosen such that there is a 4-dimensional
fixed subspace. This selects the spacetime brane, which is a solution of the IKKT model as we explained above.
 The brane solutions of the orbifolded model respect the $\Z_3$ symmetry as long as they 
do not move off the fixed plane. In order for a brane to be able to move off this plane, one has to 
add its image branes under the $\Z_3$ symmetry, which are two more branes. This extends the size of the 
matrices by three times and the gauge group at the origin appears initially to be $U(3N)$.

Up to here we have only considered the $\Z_3$ action as far as the $SO(10)$ symmetry is concerned. 
However, the discrete group acts on the matrices with respect 
to the $U(3N)$ gauge symmetry as well. Let us denote the action of $\Z_3$ on the $U(3N)$ gauge sector as $\gamma \in U(3N)$. 
This is necessarily an action through the adjoint representation, i.e. it is given by
\bea 
\g( X^{\m})&=& \g X^{\m}\g^{-1}, \\ 
\g(Z_i) &=& \g Z_i\g^{-1}.
\eea A convenient representation for $\g$ is
\be \g= \left(\begin{array}{lll}
\one_N & 0 & 0 \\ 
0 & \o\one_N & 0 \\ 
0 & 0 & \o^2\one_N
\end{array}\right)\equiv V\otimes \one_N,\ee where we have defined the $3\times 3$ matrix
\be  V=\left(\begin{array}{lll}
1 & 0 & 0 \\ 
0 & \o & 0 \\ 
0 & 0 & \o^2
\end{array}\right), \ee which satisfies $V^3=\one_3$.

In the same spirit, the action of $\Z_3$ on the spinor matrices may be defined. In particular, the corresponding actions are
\be\label{fermionso6} \z( \Psi) = \o^{b_i}\Psi \ee for the $SO(6)$ symmetry and 
\be \g( \Psi)= \g\Psi\g^{-1} \ee for the gauge summetry. In eq. (\ref{fermionso6}) the parameters $b$ are four integers, one for each of the 4-spinors, which are specifically related to the $a_i$ as 
$$
b_0 =  \sum a_i, \quad 
b_i=a_i-a_{i+1}-a_{i+2}.
$$

The orbifold matrix model is defined as the invariant sector of the IKKT model
under the full action of $\Z_3$ \cite{Aoki:2002jt}. In order to be more specific, let us consider the matrices $Z_i$ and write them in a $3\times 3$ block form,
\be Z_i=
\left(\begin{array}{lll}
z_i^{(11)} & z_i^{(12)} & z_i^{(13)} \\ 
z_i^{(21)} & z_i^{(22)} & z_i^{(23)} \\ 
z_i^{(31)} & z_i^{(32)} & z_i^{(33)}
\end{array}\right).
\ee  Similar expressions will be considered for the matrices $X^{\m}$ and $\Psi$ with entries $x^{\m(ij)}$ and $\psi^{(ij)}$ respectively. The embedding of $\Z_3$ in $SO(10)$ that we consider here translates into the following action on the $Z_i$ matrices,
\be Z_i'=\omega^{a_i}\g Z_i\g^{-1}. \ee Invariance under the $\Z_3$ action, which is described by the condition $Z_i'=Z_i$, then implies that the following matrix equation is satisfied,
\be \left(\begin{array}{lll}\label{invariancescalar}
 z_i^{(11)} & \o z_i^{(12)} & \o^2z_i^{(13)} \\
z_i^{(21)}&\o z_i^{(22)}&\o^2z_i^{(23)}\\
z_i^{(31)}&\o z_i^{(32)}&\o^2z_i^{(33)}
\end{array}\right)=\omega^{a_i}\left(\begin{array}{lll}
 z_i^{(11)} & z_i^{(12)} & z_i^{(13)} \\
\o z_i^{(21)}&\o z_i^{(22)}&\o z_i^{(23)}\\
\o^2z_i^{(31)}&\o^2z_i^{(32)}&\o^2z_i^{(33)}
\end{array}\right).
\ee 
Clearly the embedding depends crucially on the integers $a_i$. We shall consider the following two cases:
\begin{enumerate}
 \item $a_i= (1,1,1) ~\mbox{mod}~ 3$. \\ In this case the equation (\ref{invariancescalar}) becomes,
\be \left(\begin{array}{lll}
 z_i^{(11)} & \o z_i^{(12)} & \o^2z_i^{(13)} \\
z_i^{(21)}&\o z_i^{(22)}&\o^2z_i^{(23)}\\
z_i^{(31)}&\o z_i^{(32)}&\o^2z_i^{(33)}
\end{array}\right)=\left(\begin{array}{lll}
 \o z_i^{(11)} & \o z_i^{(12)} &\o z_i^{(13)} \\
\o^2 z_i^{(21)}&\o^2 z_i^{(22)}&\o^2 z_i^{(23)}\\
z_i^{(31)}&z_i^{(32)}&z_i^{(33)}
\end{array}\right),
\ee which means that only three entries of $Z_i$ remain non-zero and in particular
\be\label{z111} Z_i=
\left(\begin{array}{lll}
0 & z_i^{(12)} & 0 \\ 
0& 0 & z_i^{(23)} \\ 
z_i^{(31)} & 0 & 0
\end{array}\right).
\ee 
Moreover, as far as the matrices $X^{\m}$ are concerned, the invariance condition is 
\be X'^{\m}=\g X^{\m}\g^{-1}, \ee and therefore we obtain
\be\label{xmu} X^{\m}=
\left(\begin{array}{lll}
x^{\m(11)} & 0 & 0 \\ 
0& x^{\m(22)} & 0 \\ 
0 & 0 & x^{\m(33)}
\end{array}\right).
\ee 
Similar results apply for the fermions. The integers $b_i$ turn out to be $b_0=0~\mbox{mod}~3$ and $b_i=(1,1,1)~\mbox{mod}~3$ and therefore 
\be \label{fmat}
\Psi_{0}=
\left(\begin{array}{lll}
\psi_0^{(11)} & 0 & 0 \\ 
0& \psi_0^{(22)} & 0 \\ 
0 & 0 & \psi_0^{(33)}
\end{array}\right), \quad \Psi_i=
\left(\begin{array}{lll}
0 & \psi_i^{(12)} & 0 \\ 
0& 0 & \psi_i^{(23)} \\ 
\psi_i^{(31)} & 0 & 0
\end{array}\right).
\ee 

According to the above analysis, there are some immediate results concerning the gauge group and the spectrum of the projected theory, as well as its amount of supersymmetry. Indeed, according to eq. (\ref{xmu}) the gauge group of the resulting theory is $U(N)^3$. Moreover, according to eqs. (\ref{z111}) and (\ref{fmat}) the bosonic and the fermionic matrices $Z_i$ and $\Psi_i$ of the theory transform under the gauge group in the following bifundamental representations,
\be\label{spectrum1} 3\cdot \big( (N,\overline N,1)+(\overline N,1,N)+(1,N,\overline N)\big). \ee As far as supersymmetry is concerned, there is only one gaugino left in the projected theory, the $\Psi_0$, and therefore the theory is ${\cal N}=1$ supersymmetric. The identical spectrum for the bosonic and fermionic matrices in eq. (\ref{spectrum1}) just confirms this fact. Therefore, the spectrum contains one vector 
supermultiplet and three chiral supermultiplets. 
\item $a_i=(1,2,0)~\mbox{mod}~3$. \\
In this case the equation (\ref{invariancescalar}) has to be split in three equations, one for each $i$. The result is that the three matrices $Z_i$ assume the following form,
\bea Z_1=\left(\begin{array}{ccc}
 0&z_1^{(12)}&0\\
0&0&z_1^{(23)}\\
z_1^{(31)}&0&0
\end{array}\right), \, Z_2=\left(\begin{array}{ccc}
 0&0&z_2^{(13)}\\
z_2^{(21)}&0&0\\
0&z_2^{(32)}&0
\end{array}\right), \, Z_3=\left(\begin{array}{ccc}
 z_3^{(11)}&0&0\\
0&z_3^{(22)}&0\\
0&0&z_3^{(33)}
\end{array}\right).
 \eea
Moreover, while for the matrices $X^{\m}$ the previous result remains unaltered, for the fermionic matrices 
we obtain the integers $b_0=0~\mbox{mod}~3$ and $b_i=(2,1,0)~\mbox{mod}~3$. This gives the result
\bea \label{fmat2}
\Psi_{0}&=&
\left(\begin{array}{lll}
\psi_0^{(11)} & 0 & 0 \\ 
0& \psi_0^{(22)} & 0 \\ 
0 & 0 & \psi_0^{(33)}
\end{array}\right), \quad \Psi_{3}=
\left(\begin{array}{lll}
\psi_3^{(11)} & 0 & 0 \\ 
0& \psi_3^{(22)} & 0 \\ 
0 & 0 & \psi_3^{(33)}
\end{array}\right), \\
\Psi_1&=&\left(\begin{array}{ccc}
 0&\psi_1^{(12)}&0\\
0&0&\psi_1^{(23)}\\
\psi_1^{(31)}&0&0
\end{array}\right), \quad
\Psi_2=
\left(\begin{array}{lll}
0 & \psi_2^{(12)} & 0 \\ 
0& 0 & \psi_2^{(23)} \\ 
\psi_2^{(31)} & 0 & 0
\end{array}\right).
\eea
Presently we end up with ${\cal N}=2$ supersymmetry, which is reminiscent of a configuration $\R^6\times \C^2/\Z_2$. The spectrum consists of a ${\cal N}=2$ vector supermultiplet and a ${\cal N}=2$ hypermultiplet.
\end{enumerate}

Out of the above two possibilities, we shall focus on the first one. The reason is that the second case leads to a ${\cal N}=2$ supersymmetric 
gauge theory on the space-time brane and therefore it cannot be phenomenologically viable due to the absence of chiral matter. On the other hand, the first case corresponds to a ${\cal N}=1$ gauge theory with chiral matter content and therefore it has a promising spectrum. 

One illustrating way to rewrite the above matrices is provided by the shift and clock matrices
\be U=\left(\begin{array}{lll}
0 & 1 & 0 \\ 
0 & 0 & 1 \\ 
1 & 0 & 0
\end{array}\right), \quad V= \left(\begin{array}{lll}
1 & 0 & 0 \\ 
0 & \o & 0 \\ 
0 & 0 & \o^2
\end{array}\right). \ee These matrices satisfy the relations
\bea UU^{\dagger}&=&1, \qquad \mbox{i.e.} \quad U^{\dagger}=U^{-1}, \\ U^2&=&U^{-1}, \eea 
and similarly for $V$. Moreover, they satisfy
\be UV = \o VU. \ee 
Having defined these auxilliary matrices, it is  straightforward to see that one can write the 
configurations compatible with the orbifold constraint  in the following way
%following expressions,
\bea\label{closedformx} X_{\m}&=& x_{\m}^I\otimes V^I, \\
 Z_i&=&z_i^I\otimes V^IU.\label{closedformz} \eea
These representations of the bosonic matrices are useful in order to determine the gauge group and the spectrum of the theory, as it was exhibited in the previous section. 
However, the interpretation of the orbifold states becomes more illuminating in a different basis where 
\be
\tilde\gamma = \left(\begin{array}{lll}
 		0 & \one_N & 0 \\
 		0 & 0 & \one_N \\
 		\one_N & 0 & 0 \\
 \end{array}\right)=\one_N\otimes U. 
\label{newbasis}
\ee
This just corresponds to an alternative choice for the representative element of the action of $\Z_3$ on the matrices. Indeed, it cubes to the unit matrix and its eigenvalues are $(1,\o,\o^2)$. 

The new choice of representative element of $\Z_3$ %does not change the results concerning the gauge group and the spectrum of the projected though. 
is of course physically equivalent, but
leads to different forms for the relevant matrices. Indeed, the new invariance conditions
\bea X^{\mu}&=&\tilde\gamma X^{\m}\tilde\gamma^{-1}, \\
	Z_i&=&\o^{a_i}\tilde\gamma Z_i\tilde\gamma^{-1},
\eea are now solved by matrices of the general form
\bea X^{\m}&=& \left(\begin{array}{lll}
 		x^{\m(11)} & x^{\m(12)} & x^{\m(13)} \\
 		x^{\m(13)} & x^{\m (11)} & x^{\m(12)} \\
 		x^{\m(12)} & x^{\m(13)} & x^{\m (11)}\\
 \end{array}\right), \\ Z_i &=& \left(\begin{array}{lll}
 		z_i^{(11)} & z_i^{(12)} & z_i^{(13)} \\
 		\o^2z_i^{(13)} & \o^2z_i^{(11)} & \o^2 z_i^{(12)} \\
 		\o z_i^{(12)} & \o z_i^{(13)} & \o z_i^{(11)} \\
 \end{array}\right).
\eea Clearly the gauge group and the spectrum are more obscure in this picture. 
Now, let us write these matrices in a form analogous to (\ref{closedformx}) and (\ref{closedformz}). For $X^{\m}$ we have
\be\label{newbasisx} X^{\m}=x^{\m(11)}\otimes \one + x^{\m(12)}\otimes U+x^{\m(13)}\otimes U^2\equiv x^{\mu \ I}\otimes U^I. \ee
Similarly, for $Z_i$ we obtain
\be\label{newbasisz} Z_i=z_i^{(11)}\otimes V^2 + z_i^{(12)}\otimes V^2U+z_i^{(13)}\otimes V^2U^2\equiv z_i^I\otimes V^2U^I, \ee
where 
\be V^2U=\left(\begin{array}{lll}
 		0 & 1 & 0 \\
 		0 & 0 & \o^2 \\
 		\o & 0 & 0 \\
 \end{array}\right), \qquad V^2U^2=\left(\begin{array}{lll}
 		0 & 0 & 1 \\
 		\o^2 & 0 & 0 \\
 		0 & \o & 0 \\
 \end{array}\right). \ee

Let us return now to the action and the equations of motion of the matrix model and discuss their form after the orbifold projection. The action of the IKKT model is given by eq. (\ref{IKKT-action}) and its equations of motion by eq. (\ref{ikkteoms}). It is not surprising that the corresponding action and equations of motion for the orbifold matrix model are typically given by the same equations, albeit restricted to the $\Z_3$-invariant sector. However, it is very instructive to rewrite the action in a more illustrative form, which takes into account the fact that the orbifold group acts differently on $X^{\m}$ and $Z_i$.

The bosonic part of the action (\ref{IKKT-action}) can be brought in the following form,
 \bea\label{actionfullbosonic} S_B= -\frac 1{4g^2}Tr\biggl(&& \hspace{-0.5cm} [X_{\mu},X_{\n}][X^{\m},X^{\n}]+2[X_{\m},Z_i][X^{\m},Z_i^{\dagger}]\nn \\
	&+&\frac 12[Z_i,Z_i^{\dagger}][Z_j,Z_j^{\dagger}]+2[Z_i^{\dagger},Z_j^{\dagger}][Z_i,Z_j]\biggl).
\eea 
The corresponding equations of motion are the following,
\bea
& [X_{\m},[X_{\m},X_{\n}]]+\frac 12\biggl([Z_i,[Z_i^{\dag},X_{\n}]]+[Z_i^{\dag},[Z_i,X_{\n}]]\biggl)=0, \nn\\
& {[}X_{\m},[X_{\m},Z_j]]  % +h.c.
% [X_{\m},[X_{\m},Z_j^{\dag}]]
+\frac 12\biggl([Z_i,[Z_i^{\dag},Z_j]]
%+[Z_i,[Z_i^{\dag},Z_j^{\dag}]]
+[Z_i^{\dag},[Z_i,Z_j]]
%+[Z_i^{\dag},[Z_i,Z_j^{\dag}]]
%+h.c.
\biggl)=0.
\eea 

Let us now discuss the solutions of these equations of motion. The simplest solutions correspond to commuting matrices $[X^a,X^b]=0$, exactly as in the IKKT model. which we do not discuss here. Instead, we shall proceed to the analysis of non-commutative solutions.
There are two types of non-commutative solutions of the orbifold matrix model, which we shall describe in the basis given by eq. (\ref{newbasis}). 

\paragraph{Higgs branch.}

The first class of solutions is given by the following 
background{\footnote{The parentheses are of course not necessary but they aid in the better comprehension of the solution.}},
\bea \label{higgsbrx} X^{\m}&=&(x^{\m}\otimes \one_n)\otimes \one_3, \\
	\label{higgsbrz} Z_i&=&(z_i\otimes\one_n)\otimes V^2, \eea
where the commutation relations of the matrices are given by
\bea \label{comrelxx} [x^{\m},x^{\n}]&=&i\theta^{\m\n}, \\
	\,[x^{\m},z_i]&=&0 \\
	\,[z_i,z_j^{\dagger}]&=& i\theta_{ij}\\
	\,[z_i,z_j]&=&\,[z_i^{\dagger},z_j^{\dagger}]=0. \eea
 This vacuum corresponds to the $I=0$ sector of the eqs. (\ref{newbasisx}) and (\ref{newbasisz}) and 
in the (non-commutative) super-Yang-Mills language it is the Higgs branch of the theory. The fluctuations around this vacuum 
lead to the corresponding effective action \cite{Aoki:2002jt}.
Since the internal 
bosonic matrices $Z_i$ acquire a vev, the original gauge symmetry $U(N)^3$ is spontaneously broken.
In fact, the $U(N)$ IKKT model is recovered for large $z_i$ \cite{Aoki:2002jt}. 
Therefore  the theory is not chiral in the Higgs branch. On the other hand, it admits brane embeddings with general geometry,
which is strongly suggestive for gravity.

\paragraph{Coulomb branch.}

A second solution is given by the background 
\bea X^{\m}&=&(x^{\m \ I}\otimes \one_n)\otimes U^I, \\
	Z_i&=&0, \eea with the only non-trivial commutation relation being identical to (\ref{comrelxx}). 
This class of solutions corresponds to the Coulomb branch of the corresponding super-Yang-Mills theory and it is 
restricted on the orbifold fixed point.
In the Coulomb branch the vevs of the $Z_i$ are zero and 
therefore the gauge symmetry $U(N)^3$ remains intact, i.e. the chiral theory is retained.
However, since the branes are restricted to the orbifold fixed plane, gravity appears to be 
decoupled. 
%Therefore, gravity is decoupled from the chiral theory. 
An obvious challenge is to reconcile these seemingly incompatible pictures of chiral gauge theory and gravity,
which will be addressed further in section \ref{sec:gravity}.

\section{Modifications and fuzzy extra dimensions}

Here we would like to reveal and study another class of solutions, which are obtained after a particular modification of the above model. These solutions will correspond to fuzzy spheres, i.e. they will involve Lie-type non-commutativity. It is well-known that such solutions can be obtained by adding some appropriate terms in the matrix model action \cite{Iso:2001mg} corresponding to type IIB fluxes as discussed by Myers \cite{Myers:1999ps}. Such terms do not spoil the covergence of the related matrix integrals 
and therefore the model remains in the same topological class as shown in \cite{Austing:2003kd}. 
Moreover, such a modification is reminiscent of the addition of the soft supersymmetry breaking sector 
in the field theoretical context as we discussed in section 2.

In the present case, the terms which will be added in order to modify the matrix model are the following,
\be V_{flux}= -2\a^2Tr(Z_i^{\dagger}Z_i)-i\a\e_{ijk}Tr([Z_i,Z_j]Z_k)+h.c. \ee 
Then, if 
\be V=\frac 12 Tr\biggl([Z_i,Z_i^{\dagger}][Z_j,Z_j^{\dagger}]\biggl)+2Tr\biggl([Z_i^{\dagger},Z_j^{\dagger}][Z_i,Z_j]\biggl),  \ee
it is straightforward to show that the full part of the action involving the internal bosons takes the following form,
\be V_{full}=V+V_{flux} = Tr(F_{ij}^{\dagger}F_{ij})+D^2, \ee where
\be F_{ij}=[Z_i,Z_j]-i\a\e_{ijk}Z_k^{\dagger} \ee 
and
\be D=[Z_i,Z_i^{\dagger}].\ee 
Then the minimum of the action is obtained when 
\be \langle  F_{ij} \rangle=0 \quad \mbox{and} \quad \langle D\rangle =0, \ee
i.e. when the following algebra
is satisfied by the vacuum values of the $Z_i$ and the $Z_i^{\dagger}$,
\bea \label{alg1}\left[Z_i^{},Z_j\right]&=&i\a\e_{ijk}Z_k^{\dagger}, \\
 \label{alg2} {[}Z_i^{\dagger},Z_j^{\dagger}]&=&i\a\e_{ijk}Z_k, \\ 
\label{alg3} {[}Z_i,Z_i^{\dagger}]&=& 0,
\eea %where $\theta_{ij}$ is antisymmetric in its indices{\footnote{Therefore $[Z_i,Z_i^{\dagger}]=0$ as it should.}}. 
%Moreover, for the solutions that we  consider in the present work we set $\theta_{ij}=0$. 
% Taking the limit $\a\rightarrow 0$ in the theory under consideration, i.e. sending the deformations to zero and recovering the initial model, the limiting form of the algebra in the vacuum is
% \bea \left[Z_i^{},Z_j\right]&=&0, \\
%  {[}Z_i^{\dagger},Z_j^{\dagger}]&=&0, \\ 
%  {[}Z_i,Z_j^{\dagger}]&=&i\theta_{ij}.\eea 
In order to determine the solutions of the matrix model under consideration, let us write down the $Z_i$ in the following suggestive form (no summation in the index $i$),
\be Z_i=Y_i\otimes f_i(U,V),\ee  with $U$ and $V$ as before.  Each $f_i(U,V)$ is a matrix-valued function of the matrices $U$ and $V$ 
and can be interpreted as a scalar function on an auxilliary fuzzy torus described by these two matrices.
We assume that $Y_i$ are hermitian matrices, i.e. $$Y_i^{\dagger}=Y_i.$$ Moreover, due to the $\Z_3$ projection and the resulting form of the $Z_i$, the functions $f_i(U,V)$ must have the following special form, 
\be f_i(U,V)=a_iU+b_iUV+c_iUV^2, \ee
where $a_i,b_i,c_i$ are complex coefficients. Therefore, the matrices and their conjugate ones have the form,
\bea Z_i &=&Y_i\otimes (a_iU+b_iUV+c_iUV^2), \\ 
	Z_i^{\dagger}&=& Y_i\otimes (a_i^*U^2+b_i^*V^2U^2+c_i^*VU^2).\eea
Clearly, due to (\ref{alg1}) and (\ref{alg2}) the complex coefficients should satisfy certain constraints. Here we shall not go further into that since it does not add to the discussion. We shall therefore just consider the simplest and most obvious solution to the forementioned constraints, which is $$a_i=1, \quad b_i=c_i=0\quad \mbox{for every}\quad i=1,2,3.$$
Furthermore, according to the commutation relations between the $Z_i$ in the vacuum, it is also true that
\bea [Y_i,Y_j]&=&i\a\e_{ijk}Y_k,\\ Y_iY_i &=& R^2.   \eea 
% The matrices $Z_i$ and $Z_i^{\dagger}$ are independent degrees of freedom. Therefore, according to the above, the geometric interpretation of the vacuum is the following. As far as the first factor of the tensor product notation we adopted is concerned, it clearly corresponds to a fuzzy sphere for each set of independent degrees of freedom. Therefore the intermediate interpretation is that of a product of two fuzzy spheres, $S_F^2\times S_F^2$. 
% 
% However, there is more than that in the geometry of the vacuum and it is captured by the second factor of the tensor product. 
% This factor ''twists`` the above fuzzy spheres, in the sense that it transforms a diagonal configuration into a certain off-diagonal one. 
% We can therefore call this configuration a pair of twisted fuzzy spheres. 
Let us now turn back to the full matrix model and combine the basic brane solution we described with the twisted $S_F^2$ solution 
for the internal bosonic degrees of freedom.
The full bosonic sector of the matrix model, which combines the external and the internal matrices considered above, may be written as 
\bea\label{actionfullbosonic2} S_B= -\frac 1{4g^2}Tr\biggl(&& \hspace{-0.5cm} [X_{\mu},X_{\n}][X^{\m},X^{\n}]+2[X_{\m},Z_i][X^{\m},Z_i^{\dagger}]\nn \\
	&+&\frac 12[Z_i,Z_i^{\dagger}][Z_j,Z_j^{\dagger}]+2[Z_i^{\dagger},Z_j^{\dagger}][Z_i,Z_j]\nn\\
	&-&2\a^2Z_i^{\dagger}Z_i-i\a\e_{ijk}[Z_i,Z_j]Z_k+h.c.\biggl).\eea 
% The corresponding equations of motion are the following,
% \be ...\ee 

It is now straightforward to combine the solutions we found previously into the following solution for the full model:
\bea [ X_\mu, X_\nu] &=& i \theta_{\mu\nu}  \nn \\
	{[}Z_i^{},Z_j]&=&i\a\e_{ijk}Z_k^{\dagger},\nn \\
	{[}Z_i^{\dagger},Z_j^{\dagger}]&=&i\a\e_{ijk}Z_k, \nn\\ 
         {[}Z_i,Z_i^{\dagger}]&=& 0, \nn\\
%	 {[}Z_i,Z_j^{\dagger}]&=&i\theta_{ij},\nn\\
	{[}X_{\m},Z_i]&=&0\nn\\
	{[}X_{\m},Z_i^{\dagger}]&=&0 .
\eea 
We can immediately write down the vacuum solutions for the different matrices,
\bea X_{\m}&=&Y_{\m}\otimes \one, \\
	Z_i&=&Y_{i}\otimes U,\\
	Z_i^{\dagger}&=&Y_i\otimes U^2.
\label{twisted-solution}
\eea 
These are completely analogous to the solutions \eq{solution1} of the chiral gauge theory considered in section \ref{sec:chiral-gauge}.
In particular, they may lead to an extension of the MSSM as discussed in \cite{Chatzistavrakidis:2010xi}, and therefore represent phenomenologically
interesting solutions of the orbifold matrix model.

Let us now consider perturbations around the above vacuum solution. This task will determine the effective action on the resulting geometry. 
The relevant expansions are the following,
\bea \d X_{\m}&=&A_{\m}^I\otimes V^I, \\
	\d Z_i&=&\f_i^I\otimes UV^{I}, \\ 
	\d Z_i^{\dagger}&=& \f_i^{{\dagger}I}\otimes V^{-I}U^2.,
\label{bosonic-fluctuations}
\eea  
where summation is implied for the index $I$. Note that for the fields ($A_{\mu},\f_i,\f_i^{\dagger}$) $I$ is a labelling index, while for the matrices $U,V$ it is a power.

Let us now insert these expansions in the action (\ref{actionfullbosonic}). After some algebra and some cancellations we obtain the following result for the quadratic terms in the fields,
\bea\label{bosoniceffactionquad} S_B^{(2)}= -\frac 3{4g^2}Tr\biggl\{&&\hspace{-0.5cm} \biggl([Y_{\m},A_{\n}^I]-[Y_{\n},A_{\m}^{-I}]\biggl)^2
+2i\theta_{\m\n}A_{\m}^IA_{\n}^{-I}\nn\\&+&2(\o^I+\o^{-I})Y_iA_{\m}^IY_iA_{\m}^{-I}-2Y_i^2A_{\m}^IA_{\m}^{-I} \nn\\ 
&+&2[Y_{\m},\f_i^I]A_{\m}^{-I}Y_i-2\o^I[Y_{\m},\f_i^I]Y_iA_{\m}^{-I}\nn\\
%&+&2[Y_i,Y_{\m}]\f_i^IA_{\m}^{-I}+2\o^I[Y_{\m},Y_i]A_{\m}^{-I}\f_i^I\nn\\
&+&2\o^{-I}[Y_{\m},\f_i^{\dagger I}]A_{\m}^IY_i-2[Y_{\m},\f_i^{\dagger I}]Y_iA_{\m}^I\nn\\
%&+&2[Y_{\m},Y_i]A_{\m}^I\f_i^{\dagger I}-2\o^{-I}[Y_{\m},Y_i]\f_i^{\dagger I}A_{\m}^I\nn\\
&-&2\a^2\f_i^{\dagger I}\f_i^I+2[Y_{\m},\f_i^{\dagger I}][Y_{\m},\f_i^I]\nn\\
&+&Y_iY_j(\f_j^{\dagger I}\f_i^I+\f_j^I\f_i^{\dagger I}+4\f_i^{\dagger I}\f_j^I+4\f_i^I\f_j^{\dagger I})
-4Y_i^2(\f_j^I\f_j^{\dagger I}+\f_j^{\dagger I}\f_j^I)\nn\\
&-&\o^{-I}Y_i\f_i^{\dagger I}Y_j\f_j^J-\o^IY_i\f_j^IY_j\f_i^{\dagger I}\nn\\&-&4\o^IY_i\f_j^{\dagger I}Y_j\f_i^I-4\o^{-I}Y_i\f_i^IY_j\f_j^{\dagger I}
+4(\o^I+\o^{-I})Y_i\f_j^{\dagger I}Y_i\f_j^I\nn\\
&+&\o^{-I}(2i\a\e_{ijk}Y_k-Y_iY_j)\f_j^{I}\f_i^{-I}\nn\\
&+&\frac 12 Y_i\f_j^{-I}Y_j\f_i^{I}+\frac 12 Y_i\f_i^{I}Y_j\f_j^{-I}+h.c.
\biggl\}.\eea

\subsection{Relation with gravity and discussion}
\label{sec:gravity}

As explained in section 3, the effective metric on brane solutions $\cM \subset\R^{10}$
of the matrix model is governed by the metric $G^{\mu\nu}$ \eq{eff-metric}, which involves contributions from 
the brane embedding, as well as from the Poisson structure. 
In the orbifold matrix model, we focused primarily on branes $M^4 \times K \subset \R^{10}$,
where the embedding of the 4-dimensional brane is restricted to the 
orbifold fixed plane $\R^4 \subset \R^{10}$. 
This restriction may be relaxed in the Higgs branch upon introducing mirror images under $\Z_3$, 
however then chirality is lost.
Therefore the embedding degrees of freedom are constrained
on a chiral background, which appears to pose a problem for gravity.

To shed new light on this issue, we note that the recovery of the IKKT model from the 
orbifold AIS model in the Higgs branch as discussed in \cite{Aoki:2002jt}
is  very similar to an interesting observation by Ibanez  \cite{Ibanez:1998xn}. 
In that paper, it was pointed out that   
a certain $\Z_3$ orientifold model obtained from IIB string theory gives rise to a 
 $N = 1$, $SU(N)^3$ chiral theory, which is continuously connected to a model with $N = 4$ global supersymmetry
by giving a vev to all the fields. Then the $SU(N)^3$ is broken to the diagonal $SU(N)$. 
The spectrum includes 3 adjoint chiral superfields under $SU(N)$ with the adequate Yukawa couplings, such that  
the massless $SU(N)$ charged spectrum has global $N = 4$ supersymmetry (as well as a number of $SU(N)$ singlet chiral 
$N = 1$ supermultiplets that do not fall into $N = 4$ supermultiplets). Therefore in a sense the $SU(N)^3$ charged sector of the theory has 
$N = 4$ global supersymmetry and it is not chiral, however it bocomes chiral and only $N = 1$ supersymmetric in some points of the moduli space. 
For momenta smaller than the size of the vevs one has a subsector of the theory with $N = 4$ supersymmetry, and for momenta higher than 
those of the vevs the theory loses this extended supersymmetry and has only $N = 1$.
This is completely analogous to what happens in the orbifold matrix model, which is non-chiral
in the Higgs branch (hence at low energies) and emergent gravity is expected to arise, while the chiral $N=1$ theory is recovered
at a special point in the moduli space.  %Therefore in different points of moduli space we can find various aspects of the theory. 

There may also be a different way to recover gravity even for the chiral vacua 
with the brane sitting on the orbifold fixed point, following the mechanism 
proposed in \cite{Steinacker:2012ra}.
It turns out that for compactifications of the type $M^4 \times K \subset \R^{10}$
with $[X^\mu,Y^i] \neq 0$, a form of 4-dimensional gravity can arises from the moduli degrees of freedom of 
$K$, which are transmitted to the non-compact space-time $M^4$ via the 
Poisson structure $[X^\mu,Y^i] \sim i \theta^{\mu i} \neq 0$.
Such a non-vanishing mixing between space-time and internal coordinates is naturally
obtained by giving the internal dimensions non-vanishing angular momentum,
rotating along space-time.
In the orbifold case,  such 
spacetime-dependent twisted fuzzy extra dimensions can be obtained by modifying \eq{twisted-solution} as follows 
\bea
 Z_i =e^{i k_\mu X^\mu}\, Y^i  \otimes U,  % \qquad k_i\cdot X \equiv k_{i,\mu} X^\mu
\eea
where $Y^i$ is a hermitian generator of a fuzzy sphere.
This is clearly compatible with the orbifold constraints,
and satisfies the algebra
\bea
[Z_i,Z_j] &=& i \a(X) \, \e_{ijk} \, Z_k^\dagger , \qquad \a(X) = \a_0\, e^{i k_\mu X^\mu} .
\eea
Note that in the case of Minkowski signature,
 these can be solutions of the orbifold matrix model for suitable time-like $k_\mu$,
even without adding cubic 
or quadratic potential in the matrix model.
More generally, one could take any solution of  type $M^4 \times K$ of the IKKT mode, and 
obtain a solution of the orbifold model by adding mirror images under $\Z^3$.
The moduli of the compactification translate into a gravitational degrees of freedom via $\theta^{\mu i}$,
contributing to gravity  on the brane; we refer to \cite{Steinacker:2012ra} for a more detailed discussion.
Although
for the present simple compactification the resulting gravity may not be 
realistic, it might become realistic for more sophisticated compactifications.
The main point is that there exists a mechanism for brane gravity within
Yang-Mills matrix models which does carry over to the case of orbifold matrix model,
however a more detailed investigation is required.

\paragraph{Acknowledgments.}

%The pleasant and stimulating atmosphere at the 2010 Corfu Summer Institute of 
%elementary particle physics provided an opportunity for exchange of ideas and discussions
This work was partially supported by the SFB-Transregio TR33
"The Dark Universe" (Deutsche Forschungsgemeinschaft), the European Union 7th network program "Unification in the LHC era" (PITN-GA-2009-237920) and by the NTUA's programme supporting basic research 
PEVE 2009 and 2010.
% and the European Union's ITN programme "UNILHC" PITN-GA-2009-237920.
 The work of HS was supported in part by the Austrian Science Fund (FWF) under
the contract P21610-N16, and in part by a CCNY/Lehman CUNY collaborative grant.

%\section{Conclusions}

\end{document}